# Wisdom of collaborators: a peer-review approach to performance appraisal

Sofia Dokuka[1*], Ivan Zaikin[2,3], Kate Furman[2], Maksim Tsvetovat[4] and Alex Furman[2]

[1] Institute of Education, HSE University, Moscow, Russia
[2] Invitae Corporation, San Francisco, USA
[3] School of Computer Science & Robotics, Tomsk Polytechnic University, Tomsk, Russia
[4] The George Washington University, Washington, DC, USA

* Corresponding author email: sdokuka@hse.ru

**Abstract**

**Individual performance and reputation within a company are major factors that influence wage distribution, promotion and firing. Due to the complexity and collaborative nature of contemporary business processes, the evaluation of individual impact in the majority of organizations is an ambiguous and non-trivial task. Existing performance appraisal approaches are often affected by individuals' biased judgements, and organizations are dissatisfied with the results of evaluations. We assert that employees can provide accurate measurement of their peer performance in a complex collaborative environment. We propose a novel metric, the Peer Rank Score (PRS), that evaluates individual reputations and the non-quantifiable individual impact. PRS is based on pairwise comparisons of employees. We show high robustness of the algorithm on simulations and empirically validate it for a genetic testing company on more than one thousand employees using peer reviews over the course of three years.**

*Keywords: performance management, performance appraisal, individual performance, organizations, peer review, wisdom of crowds*

The performance and productivity of an individual in an organization is a key factor in employee compensation, development, promotion and firing[1]. In the modern transforming market economy, the systematic performance appraisal is a part of the strategic integration of human resources (HR) and business processes[2]. The majority of companies conduct employee performance evaluations at least annually[3]. However, this area has been one of the most troubling and costly in HR management for decades[4]; employees and management express considerable dissatisfaction with their appraisal mechanisms[1,3,5]. The majority of managers (up to 90%) feel that they fail to obtain accurate performance information[6], while most employees characterize performance appraisal as frustrating, too bureaucratic and often irrelevant[1]. Overall, many organizations (even giants such as Accenture, Deloitte and Microsoft) have reconsidered the use of performance appraisal[7]. Another problematic issue with the performance appraisal is the scarcity of this topic in broad scientific discussions, which can result in fast decisions made by management without expert consideration and external validation.

In today's workplace, many employees cannot be quantitatively evaluated due to the complex and collaborative nature of their work. Although managers and HR specialists can often identify the best and worst employees, they often stumble in the case of middle-performing employees and make evaluation errors[8]. They can also manipulate performance ratings by over- or underrating their colleagues[3], which results in a systematic bias.

Here we argue that the most accurate information on employee performance can be obtained directly from the company staff. We propose a mechanism for individual performance assessment that is based on peer review of the employee's close collaborators who have almost full information about her/his impact. Evaluations from direct collaborators are crucial, because reviewers observe the employee's performance, which is essential for rating quality[3,9,10]. Adler et al.[1] outline that reviews from multiple raters can diminish or even eliminate the bias. Empirical evidence shows that collective intelligence judgements (the so-called "wisdom of crowds") can be more accurate than estimates of individual experts[11–14] due to the absence of factors such as social influence and social desirability.

Peer Rank Score (PRS) serves as a proxy for individual impact and reputation in innovative organizations with cross-functional responsibilities and evolving duties. PRS is evaluated using the data from the employees' relative comparisons, which can be viewed as a series of pairwise comparisons. Similar to Elo[15] and TrueSkill[16], the PRS rating system is dynamically self-correcting. The idea of PRS is that employees who work together can correctly and accurately evaluate the impact of their peers.

The PRS evaluation needs a special data collection mechanism. Each employee regularly gives feedback on their colleagues' performance and simultaneously ranks them on a 2D-grid (which is similar to a 360-degree methodology). The horizontal axis represents "*teamwork*", which means activity and productivity in collaboration (a grid example is on Fig. 1). The vertical axis represents "*skill*," which means professional qualifications and knowledge of the employee. Together, these two parameters capture the major part of the individual's performance in an organization.

The PRS is an iterative algorithm based on pairwise comparisons. In the peer evaluation, the reviewer ranks their colleagues on the grid. Each evaluation of *n* employees by one reviewer is considered as a set of pairwise comparisons. The number of pairs compared is equal to the number of all possible combinations (Equation 1):

$$N = \frac{n!}{2!(n-2)!}$$ (Equation 1)

where *N* is the number of pairwise comparisons, and *n* is the number of compared peers. Thus, if the reviewer evaluates five peers, the 2D-grid creates data on 10 unique pairwise comparisons.

The outcome of each pairwise comparison is a win for the employee who receives a higher score during the evaluation procedure and a loss for their counteragent. For example, if **A** is evaluated higher than **B**, **A** receives points, while **B** loses the same number of points. Employees are reviewed on two scales: teamwork and skills, so that we can evaluate their reputation on both dimensions, as well as on the aggregate level. For the aggregate level, we propose the hypotenuse (Equation 2).

$$aggregate = \sqrt{teamwork^2 + skills^2} \quad \text{(Equation 2)}$$

At the beginning of the algorithm, each employee has zero PRS. After each evaluation, s/he is attributed the sum of the points (increments) that were lost or gained during the comparisons of this evaluation (Equation 3).

$$PRS(t) = PRS(t-1) + \sum increments(t) \quad \text{(Equation 3)}$$

The increment is a function of three factors: 1) the Reviewer Score (RS), which reflects the reputation of the person evaluating (*reviewer*); 2) the Expectation Score (ES), which is the expectation of such a comparison; 3) the Score Spread (SS), which is the spread in scores between the compared employees in this particular set of comparisons (Equation 4). The increment can also be artificially restricted by the maximum value.

$$increment = \sqrt[3]{RS * ES * SS} \quad \text{(Equation 4)}$$

*Reviewer score*

Individual performance varies and we are more likely to rely on persons with higher expertise and better reputation or, in other words, on employees with a high ranking position in the PRS distribution. The reviewer score, which varies continuously on a scale from 1 to 4, equals 4 for the reviewer with the highest PRS, and 1 for the reviewer with the lowest PRS (Equation 5),

$$RS = \frac{PRS_{Reviewer} - PRS_{Min}}{PRS_{Max} - PRS_{Min}} * 3 + 1 \quad \text{(Equation 5)}$$

where $PRS_{Max}$ and $PRS_{Min}$ are the maximum and minimum PRS values in the overall distribution (in the whole company), and $PRS_{Reviewer}$ is the PRS value of the employee who is doing the reviewing in this particular comparison.

*Expectation score*

The result of a pairwise comparison can be predictable or unpredictable. If we compare two employees, **A** and **B,** with a similar PRS, the probability that **A** wins is 0.5 (Equation 6.1). While in the case of the comparison of peers **C** and **D** with a clear dominance of **C** over **D**, we expect the victory of **C** with a high probability. Thus, if **C** wins, the pairwise comparison brings no surprise and the current PRS of the employees looks correct for these two particular persons. If the outcome is predictable, we expect little changes and a low expectation score. In the case of $PRS_{los} < PRS_{win}$ the expectation score is calculated using Equation 6.2. The maximum of the expectation score is 0.5, which means that there is no difference between the PRS scores of the winning and losing employees. The minimum value of the expectation score is 0.01, which means that these employees are far apart in the distribution.

However, if **D** beats **C**, the pairwise comparison outcome puts into question the current PRS structure and should seriously impact the PRS distribution and reallocate the employees in the ranking based on the new information. If the outcome is unpredictable ($PRS_{los} > PRS_{win}$), then the expectation score is calculated using Equation 6.3. It varies from 0.5 (in case of no difference in PRS between the winning and the losing employees) to 1 (in case of a large difference between evaluated peers).

$$0.5 \quad \text{(Equation 6.1)}$$

$$\max(0.01; 0.5 - \frac{PRS_{win} - PRS_{los}}{PRS_{range}}) \quad \text{(Equation 6.2)}$$

$$\min(0.5 + \frac{PRS_{win} - PRS_{los}}{PRS_{range}}; 1) \quad \text{(Equation 6.3)}$$

**Score spread**

Each employee has an individual ranking patterns. Some of them put their peers' marks close to each other on the grid, while others place them far apart. This specificity in ranking can result in biased estimates and shows the necessity of normalizing the scores received by each particular reviewer. We consider the position of the two particular evaluated peers in the context of all the evaluated colleagues in a particular review and use the normalized scores for the score spread calculation (Equation 7),

$$SS = \frac{Score_1 - Score_2}{Score_{Max} - Score_{Min}} \quad \text{(Equation 7)}$$

where $Score_1$ and $Score_2$ are the scores of the two compared peers, $Score_{max}$ is the maximum score in this particular set of the pairwise comparisons by this reviewer, and $Score_{min}$ is the minimum score, respectively. Due to the filtering mechanisms, the score spread varies from 0.05 to 1. The lower value (0.05) means that employee scores are close to each other, while the higher value (1) shows that the employees' performances are different in this particular comparison set.

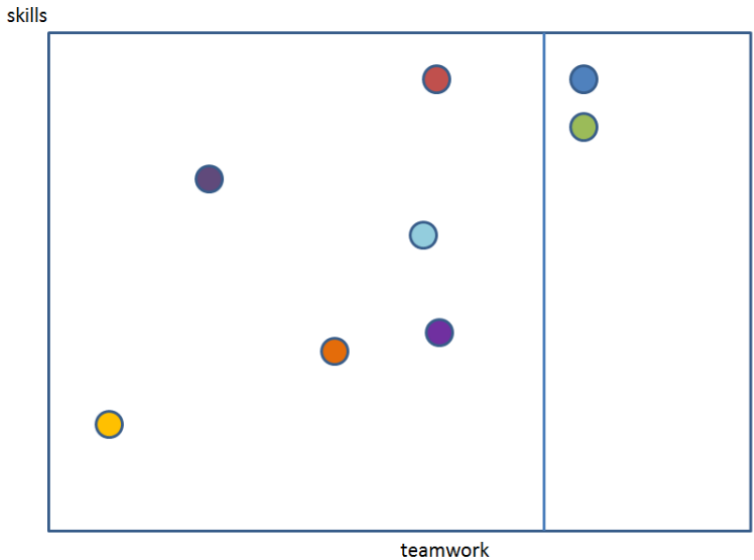

**Fig 1. 2D-grid for Peer Rank Score data gathering. The dots of the different colors represent employees. The area on the right for non-reviews.**

**Simulations**

We estimated the performance of the model with simulations of organizations of differing sizes with the randomly assigned performance and network of collaboration between the employees. We assume that each individual knows (to some extent) the performance of their collaborators. We then create the evaluation network and compute the PRS for each employee and compare it with performance.

*Step 1. Communication network formation*

Using the Erdős-Renyi algorithm[17], we construct a weighted random network with a fixed size of the organization ($N$) and the density of the employee communication connections ($p$). For each employee, we select *m* peers who were in closest connection with that employee in the simulated communication network.

*Step 2. Performance assignment*

From the normal distribution, we randomly assign the performance level (x) for each employee. We consider this to be a measure of ideal performance of the employee to be evaluated.

*Step 3. Evaluation*

Each employee evaluated the performance ($x$, which was assigned at Step 2) of their close collaborators (who were selected at Step 1).

*Step 4. Correlation calculation*

Based on the evaluations from Step 3, we calculate the PRS for each employee. Then, we measure the Spearman correlation coefficient between the computed PRS and the assigned employee performance ($x$, Step 2).

*Step 5. Noise*

In order to evaluate the algorithm resilience, we added random noise (*noise*) to the system. Random noise means that employees can make an incorrect evaluation in the assessment of their colleagues. The level of random noise varies from 0 (a system without noise) to 1 (a system with random feedback only). In the case of random noise, the evaluation is the following (Equation 8).

$$Score_{Noise} = Score_{True}(1 - noise) + Random * noise \qquad \text{(Equation 8)}$$

*Simulation results*

We perform simulations for organizations that vary in the following parameters: the size of the organization (from 100 to 1,000 employees), the level of the noise (from 0 to 1), the number of peers evaluated simultaneously. For each organization, we performed 30 simulations. After each of the evaluation cycles, we calculate the Spearman rank correlation coefficient between the obtained PRS and performance *x*. The simulation results are on Fig. 2-3. After five evaluation rounds, the correlation between the actual performance and PRS is high (more than 0.9). Overall, the correlation between actual performance and PRS is high, even in the case of random noise.

These results support the idea that groups are able to generate accurate estimates of their peers' performance, as well as the general idea of the wisdom of crowds.

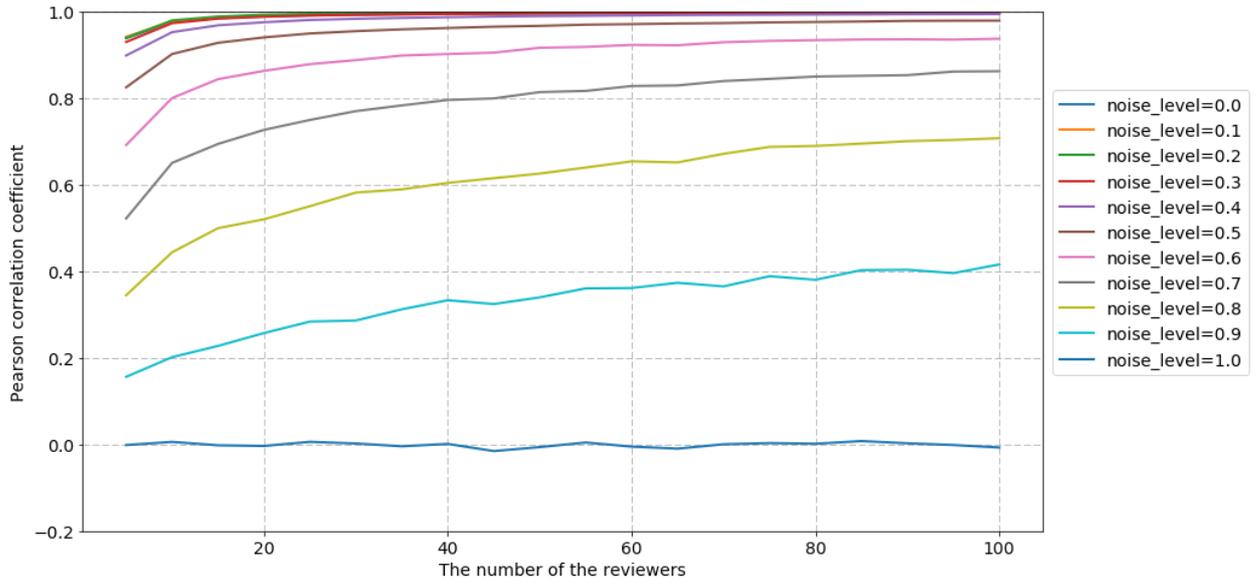

**Fig 2. Correlation between actual performance and PRS in systems with the different noise levels. The organization size is 500 employees.**

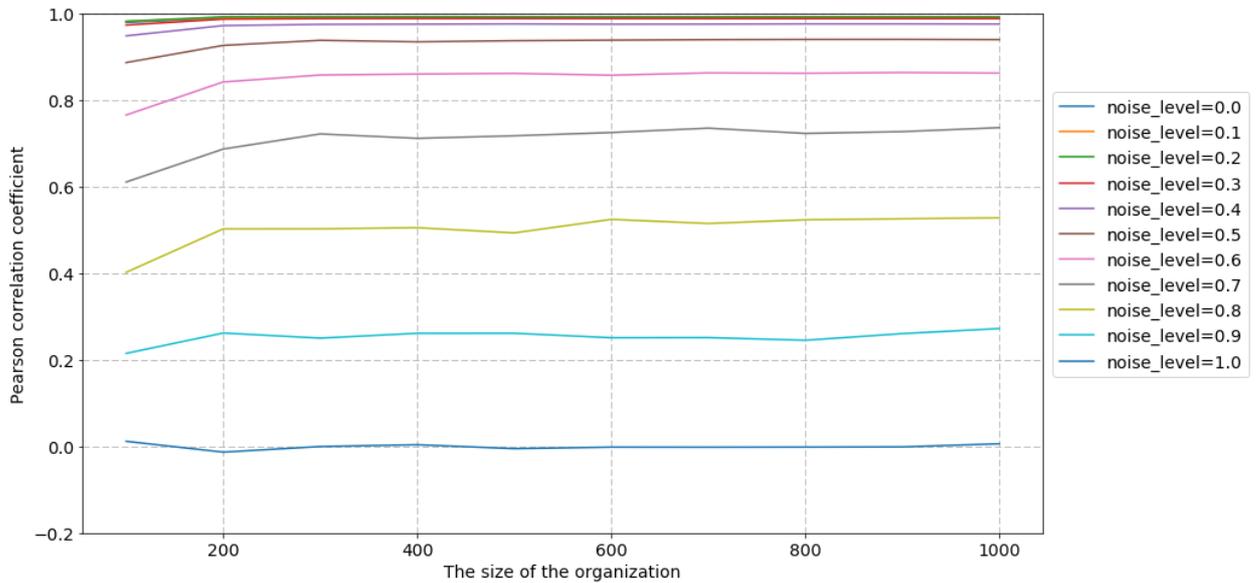

**Fig 3. Correlation between actual performance and PRS for organizations of different size. The number of reviews is 20.**

**Empirical evidence**

The PRS methodology was investigated in the case of Invitae corporation, a genetic testing company founded in 2010 and headquartered in San Francisco. Genetic tests produced by Invitae are complex and comprehensive products produced through the collaboration of many experts from different disciplines. The company has more than 1,000 employees.

The information on employee pairwise reviews has been collected staring in 2015. Employees were asked to review their colleagues in two surveys: team reviews and sampling reviews. In team review, each employee ranked up to twenty colleagues quarterly. In sampling review, conducted twice a month, each employee ranked five colleagues with whom s/he had recently communicated (based on Google calendar, email and Slack data). These two types of reviews cover both the strong and weak ties of the employees and reflect their view on their colleagues' performance. Employees have no information about how they were evaluated by others and receive no information from HR about the external assessments of their own performance. Each evaluation is carried out by the individual employee independently and anonymously, and no one can impact the final results. We expect no external impact and social influence on individual reviews. We outline the independence of employee reviews, because, according to many empirical and theoretical views, the wisdom of crowds is based on independent judgements with uncorrelated errors[12].

For validation, we study the connection between employee PRS and compensations, and PRS and positive feedback in 2017. The data on employee reviews has been collected starting in 2015, so we have more than two years of the reviews with more than 67,000 evaluations. The PRS algorithm was implemented in the HR system in April 2018; thus, there can be no influence from the PRS index on executive decisions about base salaries and grants.

For salaries, we focus on 248 employees who started their careers at Invitae before January 1, 2017 and did not leave the company in 2017. We excluded from the sample executives and top management due to the different mechanisms of their salary formation (5% of the sample). For each of the selected employees, we collected information about the overall pay they received from Invitae in 2017. The overall pay has two parts: the base salary and the performance-based equity grant (grant). The base salary, which is the fixed amount per month or per hour, depending on the type of contract, is influenced by internal and external factors, and external determinants such as market forces are crucial for the US system. The grant is not influenced by extrinsic circumstances and is mostly determined by individual efforts. This information is confidential: it is important that employees have no information about both the base salaries and the grants of their colleagues.

For each of the selected employees, we also calculated the overall amount of money earned during the year and the share the grant represents in overall compensation in the overall compensation. The PRS was calculated to the date of December 31, 2017. For each compensation type (base salary, grant, overall compensation and the share the grant represents in overall compensation), we calculated the Pearson correlation with PRS on the last day of 2017 (Fig. 4-7).

We find that there is a relatively low correlation between the base salaries and PRS ($r=0.14$, $P=0.03$). The connection between the grants and the PRS is much higher ($r=0.51$, $P<0.0001$). The overall compensation and PRS also have a medium correlation ($r=0.30$, $P<0.0001$). The highest correlation is between PRS and the share the grant represents in overall compensation ($r=0.58$, $P<0.0001$).

These results show that PRS is highly correlated with the grant and the grant part in the salary. In contrast to the baseline salary, grants are highly linked to management's perception of the employee's performance and impact. High correlation scores between PRS and grants also demonstrate the high degree of agreement in performance evaluation between the company staff overall and management.

In the second part of the validation, we trace the connection between positive feedback and PRS. Each Invitae employee can voluntary write a review of their colleague (spot review) at any time. This is a peer-driven way to reward excellent performance: employees who receive positive feedback can get bonuses such as dinner for two or a weekend getaway.

We consider a sample of 262 persons who were employed for the entire duration of 2017. For this sample, there were 164 spot review bonuses. Ninety-nine persons from the sample (38%) were nominated for bonuses in a spot review at least once, while 163 persons (62%) had no bonus nominations. The mean PRS of employees who have at least one spot review is significantly higher rather than the mean PRS of employees who have no such reviews (M=32.03 and M=11.03, respectively, P<0.0001). We also find a significant correlation between the number of spot bonuses and PRS (r=0.22, P<0.001).

These empirical evidences demonstrate the validity and effective performance of the PRS approach. The peer rank scores are highly correlated with both employee performance-based compensation, which mostly reflect the views of management and experts on the individual's impact, and positive feedback given by other employees as a reward for performance.

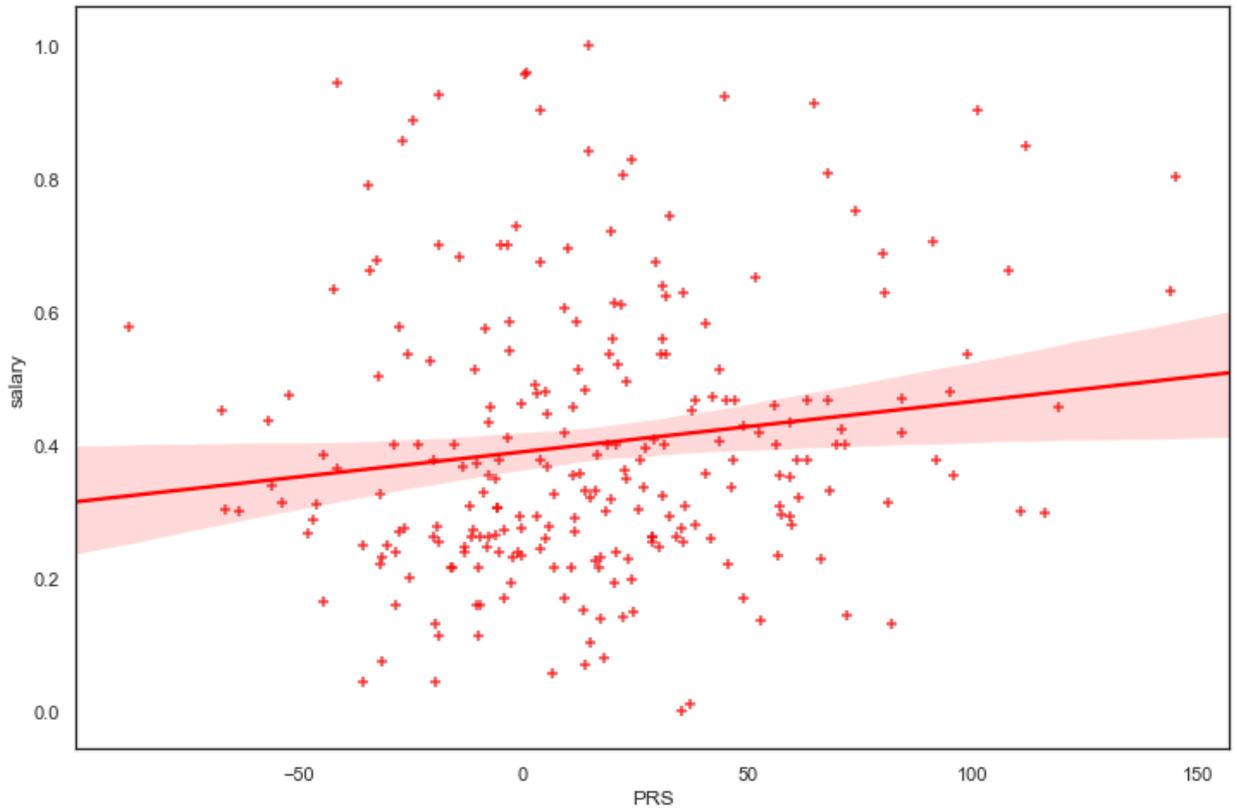

**Fig 4. Interaction between the base salary and PRS. Base salaries are normalized.**

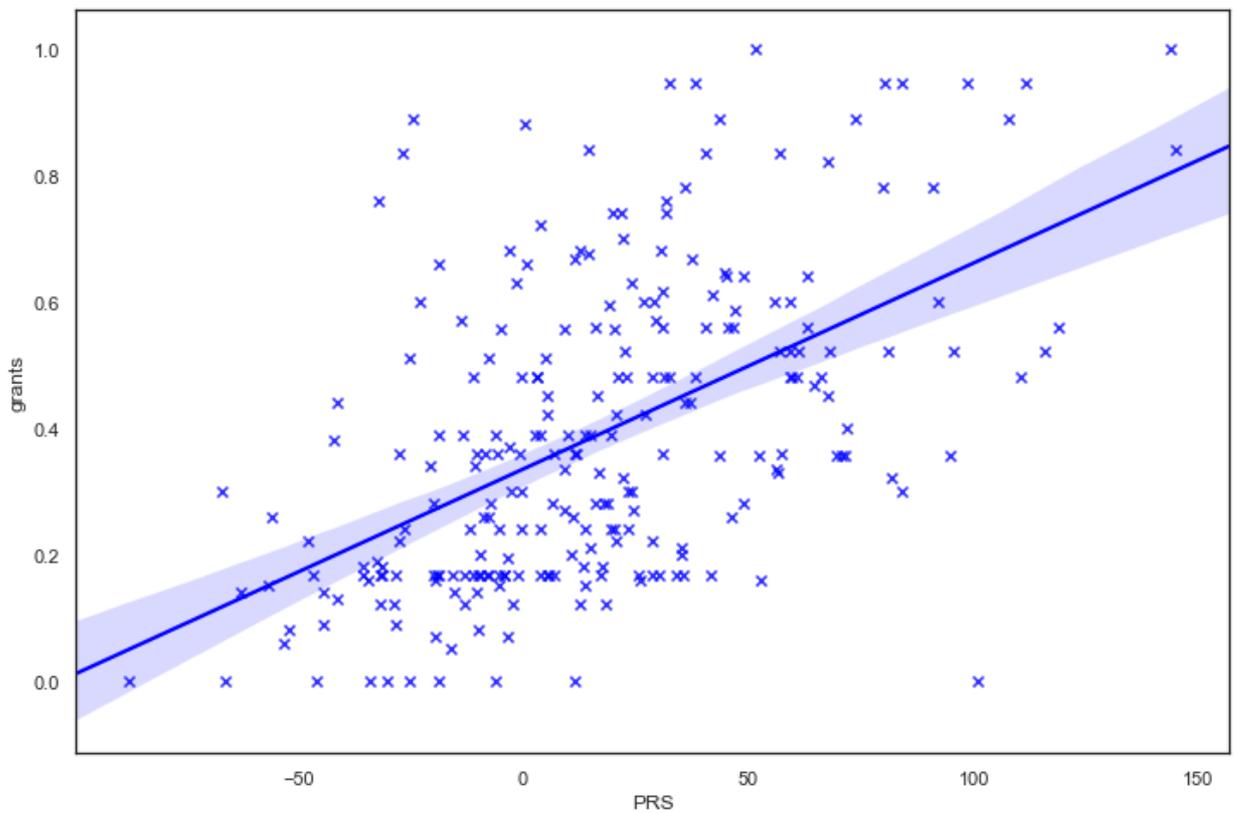

**Fig 5. Interaction between the performance-based equity grants and PRS. Grants are normalized.**

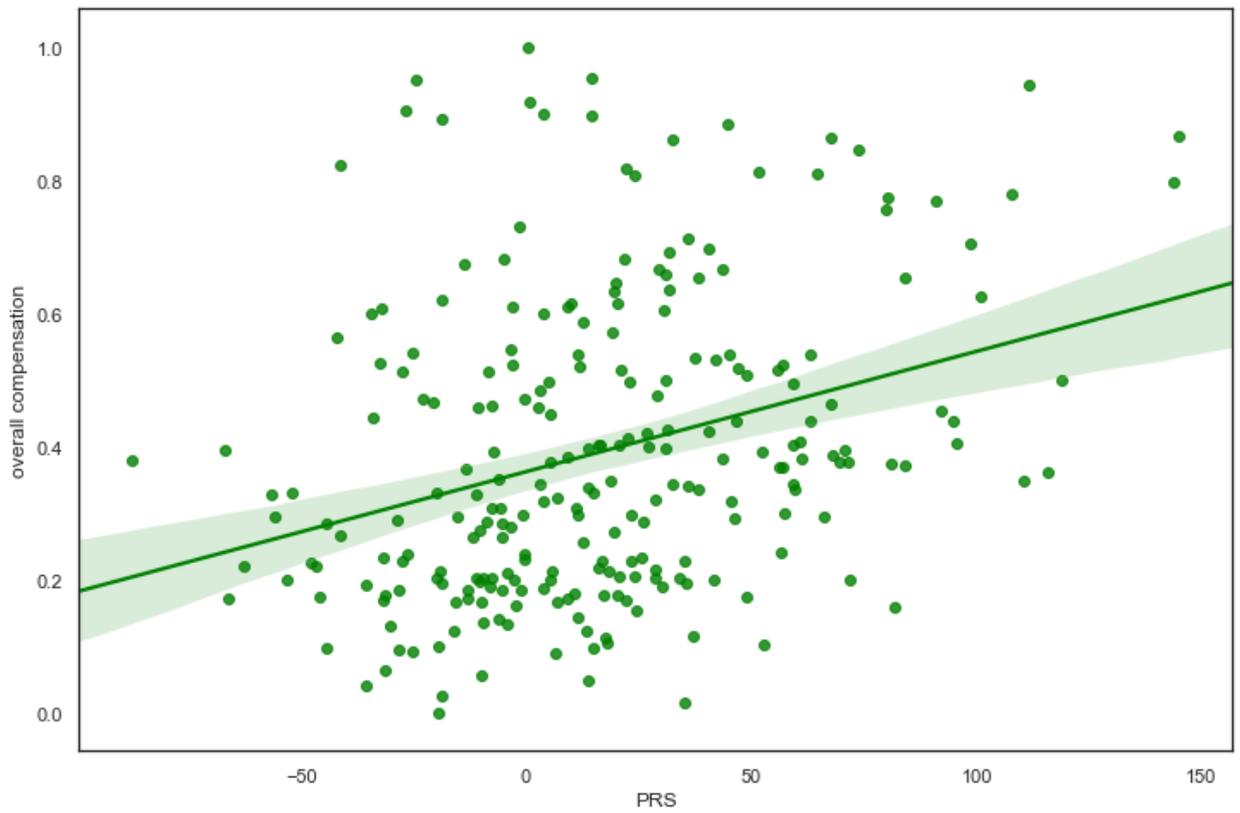

**Fig 6. Interaction between the overall compensation and PRS. Overall compensation values are normalized.**

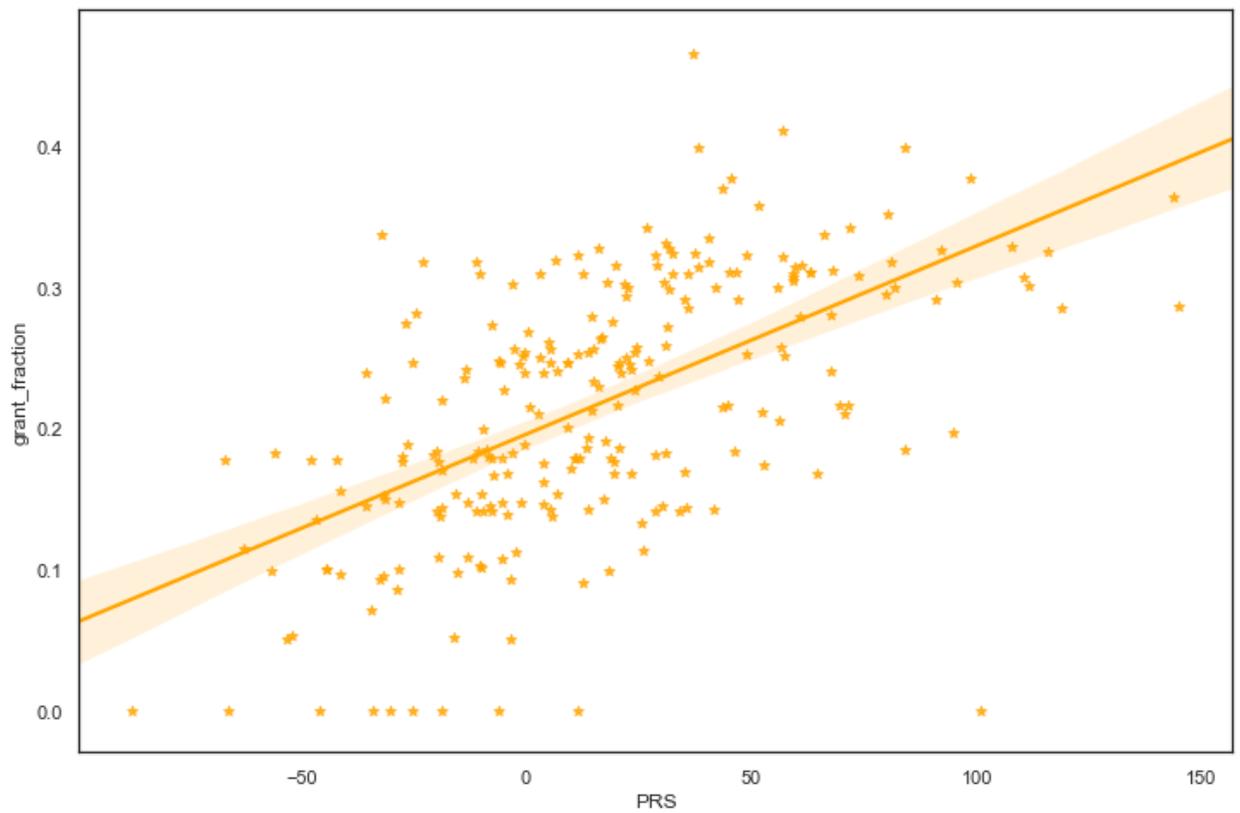

**Fig 7. Interaction the share the grant represents in overall compensation and PRS.**

**Discussion**

Performance appraisal is one of the most challenging areas in human resource management. Along with the methodological difficulties, management does not always have sufficient expertise to evaluate individual impact. Still, measurements of individual performance are necessary for making decisions regarding promotion, firing, wage distribution and, in the whole, understanding the employee dynamics and opportunities in a given organization.

In this paper, we propose an algorithm for the assessment of employee impact in the case of complex product with obscure key performance indicators. We show that employee impact can be evaluated by the employee's immediate collaborators and peers. We propose an iterative algorithm, which is based on pairwise comparisons of employees. Numerically and empirically, we show that Peer Rank Score (PRS) serves as a good measure of individual impact, which is crucial for the management purposes. In our approach, we avoid absolute measurements of the performance and consider each employee in connection with their working environment.

These results are in line with the recent agenda of the Science of Success studies, which shows that the success of the individual is the construct of group perception and recognition[18]. Our results also support the prominent idea about high performance of the wisdom of crowds[11]. We show that the expertise of even a small crowd is a powerful mechanism for the assessment of individual performance.

We also acknowledge that PRS and similar performance ratings should not be considered as an optimal outcome measurement tools in the systems with clearly defined performance metrics, e.g. call-centers or sales departments. In the case of easily determined performance with almost no collaboration, the mechanism of complex performance evaluation is not necessary. However, there is little attention given to the mechanisms of evaluation of complex tasks, which is why we emphasize the importance of such discussions in expert and scientific communities.

**Supplementary Information**

*The data collection procedure*

Every three months, each employee in the company gets a notification to review their team peers. S/he has 15 days to complete the review; if s/he does not complete the procedure in this period, s/he gets a notification to fill out the form during the following 15 days. The review procedure implies an evaluation of the set of peers on a 2D-grid. The horizontal axis represents "teamwork", which means the activity and productivity in collaboration. In the survey description, it is outlined in the following manner: "We think of teamwork as being made up of two factors: 1) Someone we trust to put the needs of the team above their own, who can be counted on to help out when needed. 2) Someone who pushes us to excel and keeps us accountable when we're not pulling our weight". The vertical axis represents "skill", which means the following: "How good the person is at their job, how much knowledge, experience, judgement they bring to the table to effectively push the team forward". These two parameters together are designed to capture the overall individual performance.

The employee sees the empty grid on the left side of the window and the subset of peers that should be evaluated below the grid. Then s/he has to drag the icons of corresponding employees into the grid according to her/his individual perception of their performance. If the reviewing employee does not know a colleague or cannot evaluate them due to other factors, s/he can move the icon of this peer out of the 2D-grid. Each review is collected once, with all the employees reviewed at the same time (an example on Fig. 1).

*Sampling formation*

In team review, an employee has to evaluate 20 peers, while in sampling review, they have to evaluate five peers.

In team review, we propose up to twenty employees to be evaluated in a single session. If the person does not have communication with at least twenty colleagues, s/he would be evaluated by fewer peers.

In contrast to team review, sampling review aims to evaluate the individual performance with a less cohesive group (so-called "weak ties"). For this purpose, we collected sampling reviews that included both teammates and peers from other teams who have been in contact (via offline meetings, emails or Slack contacts) with the reviewer in the last 90 days. In sampling review, the number of teammates that can be reviewed is limited to two.